\documentstyle[12pt]{article}

\begin{document}

\begin{center}
 {\Large \bf Factorial and Cumulant Moments in\\
  $e^{+}e^{-}\rightarrow$ Hadrons at the Z$^0$ Resonance}

\vspace {1.0cm}

{\bf The SLD Collaboration}\\

\bigskip
%
%
%
  \def\iADEL{$^{(1)}$}
  \def\iBOL{$^{(2)}$}
  \def\iBU{$^{(3)}$}
  \def\iBRUN{$^{(4)}$}
  \def\iCIT{$^{(5)}$}
  \def\iUCSB{$^{(6)}$}
  \def\iUCSC{$^{(7)}$}
  \def\iCIN{$^{(8)}$}
  \def\iCSU{$^{(9)}$}
  \def\iCOLO{$^{(10)}$}
  \def\iCOL{$^{(11)}$}
  \def\iFER{$^{(12)}$}
  \def\iFRA{$^{(13)}$}
  \def\iILL{$^{(14)}$}
  \def\iLBL{$^{(15)}$}
  \def\iMIT{$^{(16)}$}
  \def\iMASS{$^{(17)}$}
  \def\iMISS{$^{(18)}$}
  \def\iNAG{$^{(19)}$}
  \def\iOREG{$^{(20)}$}
  \def\iPAD{$^{(21)}$}
  \def\iPERU{$^{(22)}$}
  \def\iPISA{$^{(23)}$}
  \def\iRUT{$^{(24)}$}
  \def\iRAL{$^{(25)}$}
  \def\iSOGANG{$^{(26)}$}
  \def\iSLAC{$^{(27)}$}
  \def\iTENN{$^{(28)}$}
  \def\iTOH{$^{(29)}$}
  \def\iVAND{$^{(30)}$}
  \def\iWASH{$^{(31)}$}
  \def\iWISC{$^{(32)}$}
  \def\iYALE{$^{(33)}$}
  \def\dead{$^{\dag}$}
  \def\andgen{$^{(a)}$}
  \def\andper{$^{(b)}$}
%
%
\mbox{K. Abe                 \unskip,\iTOH}
\mbox{I. Abt                 \unskip,\iILL}
\mbox{C.J. Ahn               \unskip,\iSOGANG}
\mbox{T. Akagi               \unskip,\iSLAC}
\mbox{N.J. Allen             \unskip,\iBRUN}
\mbox{W.W. Ash               \unskip,\iSLAC$^\dagger$}
\mbox{D. Aston               \unskip,\iSLAC}
\mbox{K.G. Baird             \unskip,\iRUT}
\mbox{C. Baltay              \unskip,\iYALE}
\mbox{H.R. Band              \unskip,\iWISC
% ...
}
\it
\vspace {0.5cm}

%
%
  \iADEL
     Adelphi University,
     Garden City, New York 11530 \break
  \iBOL
     INFN Sezione di Bologna,
     I-40126 Bologna, Italy \break
  \iBU
     Boston University,
     Boston, Massachusetts 02215 \break
  \iBRUN
     Brunel University,
     Uxbridge, Middlesex UB8 3PH, United Kingdom \break
  \iCIT
     California Institute of Technology,
     Pasadena, California 91125 \break
  \iUCSB
     University of California at Santa Barbara,
     Santa Barbara, California 93106 \break
  \iUCSC
     University of California at Santa Cruz,
     Santa Cruz, California 95064 \break
  \iCIN
     University of Cincinnati,
     Cincinnati, Ohio 45221 \break
  \iCSU
     Colorado State University,
     Fort Collins, Colorado 80523 \break
  \iCOLO
     University of Colorado,
     Boulder, Colorado 80309 \break
  \iCOL
     Columbia University,
     New York, New York 10027 \break
  \iFER
     INFN Sezione di Ferrara and Universit\`a di Ferrara,
     I-44100 Ferrara, Italy \break
  \iFRA
     INFN  Lab. Nazionali di Frascati,
     I-00044 Frascati, Italy \break
  \iILL
     University of Illinois,
     Urbana, Illinois 61801 \break
  \iLBL
     Lawrence Berkeley Laboratory, University of California,
     Berkeley, California 94720 \break
  \iMIT
     Massachusetts Institute of Technology,
     Cambridge, Massachusetts 02139 \break
  \iMASS
     University of Massachusetts,
     Amherst, Massachusetts 01003 \break
  \iMISS
     University of Mississippi,
     University, Mississippi  38677 \break
  \iNAG
     Nagoya University,
     Chikusa-ku, Nagoya 464 Japan  \break
  \iOREG
     University of Oregon,
     Eugene, Oregon 97403 \break
  \iPAD
     INFN Sezione di Padova and Universit\`a di Padova,
     I-35100 Padova, Italy \break
  \iPERU
     INFN Sezione di Perugia and Universit\`a di Perugia,
     I-06100 Perugia, Italy \break
  \iPISA
     INFN Sezione di Pisa and Universit\`a di Pisa,
     I-56100 Pisa, Italy \break
  \iRUT
     Rutgers University,
     Piscataway, New Jersey 08855 \break
  \iRAL
     Rutherford Appleton Laboratory,
     Chilton, Didcot, Oxon OX11 0QX United Kingdom \break
  \iSOGANG
     Sogang University,
     Seoul, Korea \break
  \iSLAC
     Stanford Linear Accelerator Center, Stanford University,
     Stanford, California 94309 \break
  \iTENN
     University of Tennessee,
     Knoxville, Tennessee 37996 \break
  \iTOH
     Tohoku University,
     Sendai 980 Japan \break
  \iVAND
     Vanderbilt University,
     Nashville, Tennessee 37235 \break
  \iWASH
     University of Washington,
     Seattle, Washington 98195 \break
  \iWISC
     University of Wisconsin,
     Madison, Wisconsin 53706 \break
  \iYALE
     Yale University,
     New Haven, Connecticut 06511 \break
  \dead
     Deceased \break
  \andgen
     Also at the Universit\`a di Genova \break
  \andper
     Also at the Universit\`a di Perugia \break
\rm
%

\end{center}

\begin{center}
{\bf ABSTRACT }
\end{center}

We present the first experimental study of the ratio of cumulant to factorial
moments of the charged-particle multiplicity distribution in high-energy
particle interactions, using hadronic Z$^0$ decays collected by the SLD
experiment at SLAC.
We find that this ratio, as a function of the moment-rank $q$, decreases sharply
to a negative minimum at $q=5$, which is followed by quasi-oscillations.
These features are insensitive to experimental systematic effects and are in
qualitative agreement with expectations from next-to-next-to-leading-order
perturbative QCD.

\vfill
\eject

\section{Introduction}

One of the most fundamental observables in high-energy particle interactions is
the multiplicity of particles produced in the final state. A large body of
experimentally measured multiplicity distributions has been accumulated in a
variety of hard processes\cite{review}. The Poisson distribution (PD) does not
describe the shapes of multiplicity distributions measured in e$^+$e$^-$, pp,
and p$\bar{\rm p}$  collisions, implying non-random particle-production
mechanisms, but elucidation of the relationship between the measured shapes and
the underlying dynamics has proven to be problematic.

At present the theory of strong interactions, Quantum Chromodynamics (QCD)
\cite{qcd}, cannot be used to calculate distributions of final-state hadrons
since the mechanism of hadron formation has not been understood quantitatively.
However, perturbative QCD can be applied to calculate some properties of the
cascade of gluons radiated by the partons produced in a hard scattering process.
If there is a simple relationship between the distributions of partons and
detected final state particles, as follows for example from the ansatz of local
parton-hadron duality (LPHD) \cite{lphd}, then such calculations may be expected
to reproduce some features of experimental data. An early calculation \cite{dla}
in the leading double-logarithmic approximation (DLA) was successful in
describing the energy dependence of the average multiplicity, as well as the
energy independence of the ``KNO distribution" \cite{kno} of
$n/\!\!<\!\!n\!\!>$, the multiplicity scaled by its average value. However, the
width predicted by this calculation is much larger than that of  experimentally
observed multiplicity distributions \cite{dla}. It has been suggested that the
inclusion of higher-order terms in perturbative QCD calculations should reduce
the predicted width of the multiplicity distribution \cite{EDM,YLD3}, although
no such calculation has yet been achieved. However, the ratio of cumulant to
factorial moments has recently been proposed \cite{IMD2} as a sensitive measure
of the shape of multiplicity distributions and has been found to be calculable
in higher-order perturbative QCD.

Factorial moments have been used to characterize cascade phenomena in various
scientific fields \cite{factm}. The factorial moment of rank $q$ is defined
\cite{factm}
\begin{equation}
  F_{q} \equiv \frac{<\! n(n-1)...(n-q+1)\!>}{<\!n\!>^{q}} ,
\end{equation}
where $n$ is the particle multiplicity of an event  and
$<\!n\!>$ is the average multiplicity in the event sample.
The cumulant moments $K_q$ are related to the $F_q$ by \cite{IMD3}
\begin{equation}
  F_{q} = \sum_{m=0}^{q-1} \frac{(q-1)!}{m!(q-m-1)!} K_{q-m} F_{m},
\end{equation}
and $F_{0}\!=\!F_{1}\!=\!K_{1}\equiv 1$.
While the DLA QCD calculation predicts \cite{IMD2} that the ratio
$H_q \equiv K_{q}/F_{q}$ decreases as $q^{-2}$, the inclusion of higher
orders yields more striking behavior.
A calculation in the next-to-leading logarithm approximation (NLA) predicts
\cite{IMD1} a minimum in $H_q$ at $q\!\approx\!5$ and a positive constant
value for $q\!\gg\!5$, while the
next-to-next-to-leading logarithm approximation (NNLA) predicts \cite{IMD4}
that this minimum is negative and is followed by quasi-oscillations about
zero.  These predictions are illustrated in Fig. 1.

In a previous study \cite{LEP} $H_q$ were calculated using published
multiplicity distributions from e$^+$e$^-$ and p$\bar{\rm p}$ collisions,
and features qualitatively similar to those predicted by the NNLA
calculation were observed. This was a significant result, supporting not only
QCD at the parton level, but also the notion of LPHD. However, no account was
taken of experimental systematic effects or of the correlations, both
statistical and systematic, between values of $H_q$ at different ranks $q$. In
addition, some $H_q$ values derived from data from similar experiments were
apparently inconsistent. Furthermore, it was shown subsequently 
\cite{truncation} that the observed features could be induced by the effective
truncation of the multiplicity distribution inherent in a measurement using a
finite data sample.

In this letter we present the first experimental determination of the ratio of
cumulant to factorial moments of the charged-particle multiplicity distribution
in high-energy particle interactions, using hadronic decays of $Z^0$ bosons
produced in e$^+$e$^-$ annihilations. We study systematic effects in detail, in
particular the influence of truncation of the distribution, and investigate the
correlations between moments of 
different rank. We compare our measurements with
the predictions of perturbative QCD, and also with two widely used distributions
predicted by phenomenological models of particle production.

\section{Charged Multiplicity Analysis}

Hadronic decays of Z$^0$ bosons produced by the SLAC Linear Collider (SLC)
were collected with the SLC Large Detector (SLD) \cite{sld}.
The trigger and initial selection of hadronic events are described in
\cite{as}.
The analysis used charged tracks measured in the central drift
chamber (CDC) \cite{cdc} and vertex detector (VXD) \cite{vxd}.
A set of cuts was applied to the data to select well-measured tracks
and events well contained within the detector acceptance. Charged tracks
were required to have:
a distance of closest approach transverse to the beam axis
within 5 cm, and within 10 cm along the axis from the measured interaction
point;
a polar angle, $\theta$, with respect to the beam axis within
$|\cos\theta| < 0.8$; and
a momentum transverse to the beam axis greater than $0.15$~GeV/c.
Events were required to have:
a minimum of five such tracks;
a thrust-axis \cite{thrust} direction within $|\cos\theta_T| < 0.71$; and
a total visible energy of at least 20 GeV, which was
calculated from the selected tracks assigned the charged pion mass.
A total of 86,679 events from the 1993 to 1995 SLC/SLD runs survived
these cuts and were included in this analysis.
The efficiency for selecting hadronic events satisfying the
$|\cos\theta_T|$ cut was estimated to be above $96\%$. The background
in the selected event sample was estimated to be ($0.3 \pm 0.1$)\%,
dominated by Z$^0 \rightarrow \tau^+ \tau^-$ decays, and was subtracted
statistically from the observed multiplicity distribution.

The experimentally observed charged-particle multiplicity distribution was
corrected for effects introduced by the detector, such as geometrical
acceptance, track-reconstruction efficiency, and additional tracks from photon
conversions and particle interactions in the detector materials, as well as for
initial-state photon radiation and the effect of the cuts listed above. The
charged multiplicity of an event was defined to include all promptly produced
charged particles, as well as those produced in the decay of particles with
lifetime $< 3 \cdot 10^{-10} s$. A two-stage correction was calculated using
Monte Carlo simulated hadronic Z$^0$ decays produced by the JETSET 6.3
\cite{jetset63} event generator, subjected to a detailed simulation of the SLD
and reconstructed in the same way as the data.  Each MC event passing the
event-selection cuts yielded a number of generated tracks $n_g$ and 
a number of
observed tracks $n_o$, which were used to form the matrix
\begin{equation}
  M(n_g,n_o) = \frac{N(n_g,n_o)}
                    {N_{obs}^{MC}(n_o)}  ,
\end{equation}
where $N(n_g,n_o)$ is the number of MC events with $n_g$
generated tracks and $n_o$ observed tracks, and $N_{obs}^{MC}(n_o)$ is the
number of MC events with $n_o$ observed tracks. For each $n_o$, a sum of three
Gaussians was fitted to $M(n_g,n_o)$ and this parametrization was used in the
correction. The effects of the event-selection cuts and of initial-state
radiation were corrected using factors
\begin{equation}
  C_F(n_g) = \frac{P^{true}(n_g)}
                  {P^{sel}(n_g)} ,
\end{equation}
where $P^{true}(n_g)$ is the normalized simulated multiplicity distribution
generated without initial-state radiation and $P^{sel}(n_g)$ is the normalized
distribution for those events in the fully-simulated sample that passed the
selection cuts.

Both corrections were applied to the experimentally observed multiplicity
distribution $P^{exp}(n_o)$ to yield the corrected distribution:
\begin{equation}
  P^{cor}(n) = C_F(n)\cdot \sum_{n_o} M(n,n_o)\cdot P^{exp}(n_o),
\end{equation}
which is shown with statistical errors only in Fig. 2a. The factorial moments
$F_q$, cumulant moments $K_q$, and their ratios $H_q$ were calculated from this
distribution according to Eqs.~1 and 2.  The resulting $H_q$ up to rank $q =17$
are shown in Fig.~3 and listed in Table~1. As $q$ increases, the value of $H_q$
falls rapidly (inset of Fig.~3), reaches a negative minimum at $q\!=5$,  and
then oscillates about zero with a positive maximum at $q\!= 9$ and a second
negative minimum at $q\!=13$. The statistical and systematic errors are strongly
correlated between ranks as  we now discuss.

\section{Statistical and Systematic Errors}

Statistical errors and correlations were studied by analyzing simulated
multiplicity distributions. The $H_q$ were calculated from 10 Monte Carlo
samples of the same size as the data sample and 20 multiplicity distributions
generated according to the measured distribution. For each $H_q$ the standard
deviation in these 30 samples was taken as the statistical error, and is listed
in Table~1. In each case the $H_q$ exhibited the same behavior as those
calculated from the data, although the value of $H_{5}$ and the apparent phase
of the quasi-oscillation for $q\!\geq 8$ were found to be sensitive to
statistical fluctuations. We investigated the possibility that the observed
features might result from a statistical fluctuation by generating 10,000
multiplicity distributions according to Poisson and negative-binomial
distributions (see below) with the same mean value as our corrected multiplicity
distribution.  In no case did any sample exhibit either a minimum near $q\!=5$
or quasi-oscillations at higher $q$.

Experimental systematic effects were also investigated. An important issue is
the simulation of the track-reconstruction efficiency of the detector. The $H_q$
were found to be sensitive to the global efficiency, which was tuned in the
simulation so that our average corrected multiplicity equalled the value
measured in hadronic $Z^0$ decays \cite{dcone}.   The $H_q$ resulting from a
variation in the global efficiency of $\pm 1.7\%$, corresponding to the error on
the measured average multiplicity,  are shown in Fig.~4. There is an asymmetric
effect on the value of $H_5$ and on the apparent phase of the quasi-oscillation.
For each $q$ the difference between the $H_q$ with increased and decreased
efficiency was assigned as a symmetric systematic uncertainty.

It is important to consider the dependence of the track reconstruction
efficiency on multiplicity. Our simulated efficiency is 91.5\% for tracks
crossing at least 40 of the 80 layers of the CDC, and is independent of $n_g$
within $\pm 0.5\%$. Varying the efficiency for $n_g\!\!> 20$ by $\pm 0.5\%$
caused a change of $\pm 4\%$ in $H_5$, and negligible changes for $q>5$. This
change was assigned as a systematic uncertainty.

Variation of the form of the parametrization of the correction matrix $M$ was
found to affect mainly the amplitude of the quasi-oscillation for $q\!\geq 8$.
Application of the unparametrized version of the matrix $M(n_g,n_o)$ produced
the largest such effect, which is shown in Fig.~4. This change was
conservatively assigned as a symmetric systematic uncertainty to account for
possible mismodelling of the off-diagonal elements of the matrix. The effect on
the $H_q$ of variation of the parameters of the three-Gaussian fits to $M$
within their errors increases with increasing $q$, becoming the dominant
uncertainty for $q\geq$16.

The effects on the $H_q$ of wide variations in the criteria for track and event
selection were found to be small compared with those due to the above sources.
The effect of including values of the multiplicity distribution at $n\!=\!2$ and
$n\!=\!4$, taken from the JETSET model, in the calculation of the moments is
also small. Varying the estimated level of non-hadronic background, which
appears predominantly in the low-multiplicity bins, by $\pm 100\%$ produces a
negligible change in the $H_q$.

The uncertainties from the above systematic sources were added in quadrature to
derive a systematic error on each $H_q$, which is listed in Table 1. All of our
studies showed a clear first minimum in $H_q$ at $q=5$ followed by
quasi-oscillations for $q\!\geq$8. The value of $H_5$ has a total uncertainty of
$\pm 13\%$ that is strongly correlated with similar errors on $H_6$ and $H_7$
and with an uncertainty in the phase of the quasi-oscillation of $\mp 0.2$ units
of rank. There is an uncertainty on the amplitude of the quasi-oscillation of
$\pm 15\%$ that is essentially independent of the other errors. From these
studies   we conclude that the steep decrease in $H_q$ for $q\!<$5, the negative
minimum at $q=5$, and the quasi-oscillation about zero for $q\!\geq$8 are
well-established features of the data.

\begin{table}[tbh]
\label{tab:hq}
\begin{center}
\begin{tabular}{|r|r|c|c|}\hline
             &   \hfill $H_q$\quad  \hfill    & Statistical & Systematic \\
          $q$&  \hfill ($10^{-4}$)  \hfill &  error      & error    \\ \hline
           2 & 411.00     &  2.96       &  11.13    \\
           3 &  54.41     &  1.40       &  \phantom{1} 5.61    \\
           4 &   5.15     &  0.74       &   \phantom{1} 0.93    \\
           5 & $ -4.08$     &  0.40       &  \phantom{1}  0.51    \\
           6 & $ -3.40$     &  0.28       &   \phantom{1} 0.39    \\
           7 & $ -1.40$    &  0.20       &   \phantom{1} 0.32    \\
           8 &   0.08     &  0.14       &   \phantom{1} 0.10    \\
           9 &   0.91     &  0.12       &  \phantom{1}  0.16    \\
          10 &   0.84     &  0.10       &  \phantom{1}  0.19    \\
          11 &   0.10     &  0.08       &   \phantom{1} 0.09    \\
          12 & $ -0.66$     &  0.10       &  \phantom{1}  0.15    \\
          13 & $ -0.83$     &  0.09       &   \phantom{1} 0.21    \\
          14 & $ -0.18$     &  0.10       &  \phantom{1}  0.13    \\
          15 &   0.89     &  0.16       &   \phantom{1} 0.26    \\
          16 &   1.50     &  0.19       &   \phantom{1} 0.45    \\
          17 &   0.61     &  0.26       &   \phantom{1} 0.37    \\ \hline
\end{tabular}
\caption {Ratio of cumulant to factorial moments, $H_q$. The errors are strongly
correlated between ranks as discussed in the text.}
\end{center}
\end{table}

\section{Comparison of the $H_q$ with QCD Predictions}

We have compared these results with the qualitative predictions of perturbative
QCD discussed in Section 1.  Figure~1  shows that the DLA QCD calculation
predicts no negative values of $H_q$ and is inconsistent with the data. The NLA
and NNLA calculations predict \cite{LEP} a steep decrease in $H_q$  to a minimum
at
\[ q_{min} = \Bigl[ \frac{96\pi}{121\alpha_s(Q^2)} \Bigr]^{1/2} + \frac{1}{2} .
 \]
For $\alpha_s(M_Z^2)$ measured in $Z^0$ decays \cite{as} $q_{min}\!\approx 5$.
These features are seen in the data. For $q\!> 5$, the NLA calculation predicts
that $H_q$ increases toward a constant value, which is not consistent with the
data, whereas the NNLA calculation predicts quasi-oscillations in $H_q$ in
agreement with the data.

The moment ratios are thus seen to be a sensitive discriminator between QCD
calculations at different orders of purturbation theory. We conclude that the
$H_q$ calculated for gluons in the next-to-next-to-leading logarithm
approximation of perturbative QCD describe the shape of the observed
multiplicity distribution, whereas the available calculations at lower order do
not.

\section{Comparison with phenomenological models}

Measured multiplicity distributions have been compared extensively with the
predictions of phenomenological models. We consider two such predicted
distributions. The negative binomial distribution (NBD)
\begin{equation}
  P_{n}(\langle n\rangle,k) = C^{n}_{k+n-1}
       \left(\frac{\langle n\rangle}{\langle n\rangle + k}\right)^n
                   \left(\frac{k}{\langle n\rangle + k}\right)^k ,
\end{equation}
where $\langle n\rangle$ and $k$ are free parameters, is predicted \cite{nbdthy}
by models in which the hard interaction produces several objects, sometimes
identified with the partons in a QCD cascade, each of which decays into a number
of particles.   The log-normal distribution (LND)
\begin{equation}
  P_{n}(\mu, \sigma, c) = \int_n^{n+1} \frac{N}{n' + c}
     \exp \left(- \frac{(\ln(n'+c)-\mu)^2}{2\sigma^2} \right) dn',
\end{equation}
where $\mu$, $\sigma$, and $c$ are free parameters, is predicted \cite{lndthy}
by models in which the particles result from a scale-invariant stochastic
branching process, which might be related to the parton branchings in a QCD
cascade.

Considering statistical errors only, we performed least-squares fits of the NBD
and LND to our corrected multiplicity distribution. These fitted distributions
and their normalized residuals are shown in Figs. 2a and 2b, respectively. Both
provide reasonable descriptions of the data, with $\chi^{2}/$ndf of 68.0/24 and
30.5/23, respectively. Although the NBD has a high $\chi^{2}$ and shows
structure in the residuals in the core of the distribution, it is difficult to
exclude without a thorough understanding of the uncorrelated component of the
systematic errors. These results are in agreement with those from a previous
analysis \cite{lndexp}.

The PD and the phenomenological distributions differ markedly in their moment
structure: for the PD, $H_q=0$ for all $q$; for the fitted NBD, $H_q$ is
positive and falls as $q^{-25}$; for the fitted LND, $H_q$ falls with increasing
$q$ to a negative minimum at $q\!=6$ and then oscillates about zero. It was
recently argued \cite{truncation}~that the truncation of the large-$n$ tail of
the multiplicity distribution due to finite data-sample size could lead to
quasi-oscillations in $H_q$ similar to those observed in the data.   We
calculated $H_q$ values from the fitted distributions over the multiplicity
range observed in the data, $6 \leq \!n\! \leq 54$, and the results are
displayed in Fig. 5. The truncated PD and NBD  are found to produce features
similar to those in the data, but with much smaller amplitudes. The amplitudes
are not sensitive to the exact value of the truncation point and we conclude
that the moment ratios predicted by the PD and NDB are inconsistent with the
data. The LND predictions  are insensitive to the truncation point and show the
same qualitative features as the data.  However, the first minimum is smaller in
amplitude and is at $q\!= 6$.  The quasi-oscillation for $q\!\geq 8$ has similar
amplitude and period, and is displaced by about one unit from the data. The
moment ratios $H_q$ are thus seen to provide a sensitive test of
phenomenological models.

\section{Conclusion}

In conclusion, we have conducted the first experimental study of the ratio $H_q$
of cumulant to factorial moments of the charged-particle multiplicity
distribution in high-energy particle interactions, using hadronic $Z^0$ decays. 
We find that $H_q$ decreases sharply with increasing rank $q$ to a negative
minimum at $q=5$, followed by quasi-oscillations; we show these features to be
insensitive to statistical and experimental systematic effects.

The predictions of perturbative QCD in the next-to-next-to-leading-logarithm
approximation are in agreement with the features observed in the data,
supporting both the validity of QCD at the parton level and the notion that the
observable final state reflects the underlying parton structure. Calculations in
the leading double-logarithm and next-to-leading-logarithm approximations are
not sufficient to describe the data. The Poisson and negative binomial
distributions do not predict these features. The log-normal distribution
predicts features similar to those of the data, but does not describe the data
in detail. We conclude that the moment ratios $H_q$ of the charged-particle
multiplicity distribution provide a sensitive test both of perturbative QCD and
of phenomenological models.

\section*{Acknowledgements}

We thank the personnel of the SLAC accelerator department and the technical
staffs of our collaborating institutions for their efforts, which resulted in
the successful operation of the SLC and the SLD. We thank I. Dremin for useful
discussions.

This work was supported by U.S. Department of Energy contracts:
  DE-FG02-91ER40676 (BU),
  DE-FG03-92ER40701 (CIT),
  DE-FG03-91ER40618 (UCSB),
  DE-FG03-92ER40689 (UCSC),
  DE-FG03-93ER40788 (CSU),
  DE-FG02-91ER40672 (Colorado),
  DE-FG02-91ER40677 (Illinois),
  DE-AC03-76SF00098 (LBL),
  DE-FG02-92ER40715 (Massachusetts),
  DE-AC02-76ER03069 (MIT),
  DE-FG06-85ER40224 (Oregon),
  DE-AC03-76SF00515 (SLAC),
  DE-FG05-91ER40627 (Tennessee),
  DE-AC02-76ER00881 (Wisconsin),
  DE-FG02-92ER40704 (Yale);
  U.S. National Science Foundation grants:
  PHY-91-13428 (UCSC),
  PHY-89-21320 (Columbia),
  PHY-92-04239 (Cincinnati),
  PHY-88-17930 (Rutgers),
  PHY-88-19316 (Vanderbilt),
  PHY-92-03212 (Washington);
  the UK Science and Engineering Research Council
  (Brunel and RAL);
  the Istituto Nazionale di Fisica Nucleare of Italy
  (Bologna, Ferrara, Frascati, Pisa, Padova, Perugia);
  and the Japan-US Cooperative Research Project on High Energy Physics
  (Nagoya, Tohoku).


\vfill
\eject

\section*{Figure captions }

\begin{enumerate}
\item
Functional form of perturbative QCD predictions of the ratio $H_q$ of cumulant
to factorial moments in the leading double-logarithm (solid line),
next-to-leading-logarithm (dotted line) and next-to-next-to-leading-logarithm
(dashed line) approximations.  The vertical scale and relative normalizations
are arbitrary.

\item
a) The corrected charged-particle multiplicity distribution. The open circles at
$n\!= 2$, 4 are the predictions of the JETSET Monte Carlo. The solid and dashed
lines represent fitted negative-binomial and log-normal distributions,
respectively.  The normalized residuals are shown in b). The fits yielded
parameter values of $k=24.9$ and $\langle n\rangle = 20.7$ for the NBD and
$\mu=3.52$, $\sigma = 0.175$ and $c = 13.4$ for the LND. The errors are
statistical only.

\item
Ratio of cumulant to factorial moments, $H_{q}$, as a  function of the moment
rank $q$. The error bars are statistical and are strongly correlated between
ranks.

\item
Examples of systematic effects on  $H_{q}$. The data points show the $H_q$ with
statistical errors derived using the standard correction. The dotted (dashed)
line connects $H_q$ values derived with an increase (decrease) of 1.7\% in the
simulated track reconstruction efficiency. The solid line connects $H_q$ values
derived using the unparametrized correction matrix.

\item
Comparison of the $H_{q}$ measured in the data (dots with statistical errors)
with the predictions of  truncated Poisson (dotted line joining the values at
different $q$), negative binomial (dashed line) and log-normal (dot-dashed line)
distributions.

\end{enumerate}

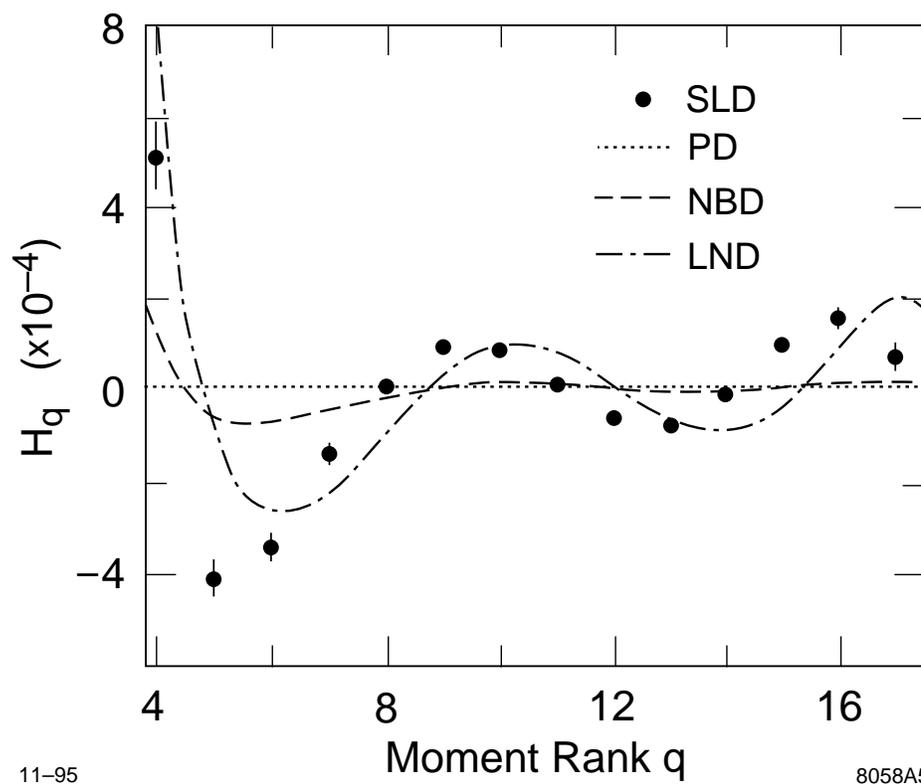
\end{document}